\documentclass[aps,amsfonts,amsmath,prd,preprint,nofootinbib]{revtex4-1}
\usepackage{epsfig,bbm,cancel}
\usepackage{slashed,xcolor}
\usepackage[normalem]{ulem} 
\usepackage{amssymb}

\newcommand{\Tr}{\text{Tr}}
\newcommand{\diag}{\text{diag}}

\begin{document}

%
\title{ BPS Submodels of The Generalized Skyrme Model and How to Find Them}

\author{Ardian Nata Atmaja$^1$}
\email{ardi002@lipi.go.id}
\author{Bobby Eka Gunara$^2$}
\email{bobby@fi.itb.ac.id (Corresponding author)}
\author{Ilham Prasetyo$^{1,3}$}
\email{ilham.prasetyo@sci.ui.ac.id}
\affiliation{$^1$ Research Center for Physics, Indonesian Institute of Sciences (LIPI),
Kompleks PUSPIPTEK Serpong, Tangerang 15310, Indonesia.}
\affiliation{$^2$Theoretical High Energy Physics and Instrumentation Research Group, Institut Teknologi Bandung,
Jl. Ganesha 10 Bandung 40132, Indonesia.}
\affiliation{$^3$Departemen Fisika, FMIPA, Universitas Indonesia, Depok 16424, Indonesia.}

\def\changenote#1{\footnote{\bf #1}}

\begin{abstract}

Using the BPS Lagrangian method we show that all known BPS submodels of the generalized Skyrme model, with a particular ansatz for the fields content, can be devided into three groups based on the (effective) number of derivative-terms in the BPS submodels. We are able to derive rigorously the Bogomolny's equations of those BPS submodels. The resulting Bogomolny's equations, along with possible constraint equations, are in general forms in which some of the known BPS submodels may contain other possible non-trivial (non-vacuum) solutions then the ones found in the literature. Furthermore, we derive some other new BPS submodels of the generalized Skyrme model for each of the groups and some of them yield new solutions.


\end{abstract}

\maketitle
\thispagestyle{empty}
\setcounter{page}{1}

\section{INTRODUCTION}

In large-$N_c$ limit, the QCD is known to be equivalent to an effective theory of mesons~\cite{tHooft:1974pnl}. Later on it was shown that baryons can be understood as solitons in this theory of mesons~\cite{Witten:1979kh}. Although the large-$N_c$ theory of mesons is not fully understood, at least at low energy this theory reduces to a nonlinear sigma model of spontaneously broken chiral symmetry~\cite{Adkins:1983ya}. Furthermore solitons of the nonlinear sigma model possess the quantum numbers of QCD baryons~\cite{Witten:1983tx}. The Skyrme model is a type of nonlinear sigma model which contains a topological soliton known as Skyrmion and it was initially proposed to model the nucleon~\cite{Skyrme:1961vq,Skyrme:1961vr,Skyrme:1962vh}. This model has been generally proposed to describe the low energy hadrons dynamics in which the primary fields are mesons and baryons appears as its solitonic excitations~\cite{Adkins:1983ya,Jackson:1983bi,Adkins:1983hy}. The Skyrmion static energy has a lower bound proportional to a topological degree $B$, which is identified as baryon number. This is known from employing the original Bogomolny method~\cite{Bogomolny:1975de,Faddeev:1976pg}. Unfortunately, the only solutions satisfying the Bogomolny equations are trivial one, $B=0$,~\cite{Manton:1987xt}. For a comprehensive review, see~\cite{Zahed:1986qz,Makhankov:1993ti}. Other interesting application of Skyrmion can be found in various studies. One notable example is magnetic skyrmions in condensed matter physics (see, e.g., \cite{PhysRevB.47.16419,Fert2013,Romming636}). This, and other studies as well, had inspired a lower dimensional model called baby Skyrme model~\cite{Piette:1994jt,Piette:1994ug,Piette:1994mh}. More recently, its BPS bound, which is saturated by its BPS solutions, had been found~\cite{Adam:2012pm}. Others also study the model that also include the Chern-Simons term~\cite{Samoilenka:2016wys}.

Bogomolny equations has extensively used in some theories especially with topological solitons. First implemented to nonabelian monopoles and dyons, the Bogomolny method~\cite{Bogomolny:1975de}, by arranging the terms to have the form of squared terms plus some boundary terms, is able to produce Bogomolny equations which are first order and satisfy the exact solutions of monopoles and dyons, or known as BPS monopoles/dyons, found by Prasad and Sommerfield~\cite{Prasad:1975kr}. Total energy of solutions of Bogomolny equations saturate its lowest energy bound, which turn out to be proportional to its topological degree. We sometimes called the Bogomolny equations that has non-trivial solutions as BPS (Bogomolny-Prasad-Sommerfield) equations.

Other than monopoles and dyons, the Nielsen-Olesen magnetic vortices~\cite{Nielsen:1973cs} are also known to have BPS equations. There are some generalizations that also has BPS equations, for instance when the form of kinetic term is similar to Dirac-Born-Infeld for vortices~\cite{Shiraishi:1990zi} and for nonabelian monopole and dyon~\cite{Nakamula:2013rfa}. Since the Bogomolny method is harder to implement to more general models, there has been many proposals for refinement, for instance the first-order formalism~\cite{Bazeia:2005tj,Bazeia:2007df}, the concept of strong necessary conditions~\cite{Sokalski:2001wk,Adam:2016ipc,Stepien:2015naq,Stepien:2016qzz}, the on-shell method~\cite{Atmaja:2014fha,Atmaja:2015lia}, and the BPS Lagrangian method~\cite{Atmaja:2015umo,Atmaja:2018ddi}.

In Skryme model, since the Bogomolny equations from the original one has no non-trivial(non-vacuum) solution, there has been proposed at least two modified models, i.e. a sextic term in first derivative and a potential~\cite{Adam:2010fg,Adam:2010ds,Adam:2015ele}, or known as BPS Skyrme model, and a quartic term in first derivative and a potential~\cite{Harland:2013rxa}. Recently, it has been found that the Skyrme model has two submodels~\cite{Adam:2017pdh}. One of them had been found to contain two subsubmodels~\cite{Stepien:2018mti}.

In this paper, we will use the BPS Lagrangian method proposed in~\cite{Atmaja:2015umo} to find Bogomolny equations in some submodels of the generalized Skryme model. This method initially had been used for some models of vortices and it had also been used for nonabelian magnetic monopoles and dyons~\cite{Prasetyo:2018yzl,Atmaja:2018cod}. 
This paper is organized as follows. In the next section, we revisit the Skyrme model just until the effective Lagrangian density from a given ansatz. The method is given in the following section with a subsection for explanations on modifications for this model. Then in the following sections, we implement the method to some known submodels and some new ones.

\section{THE SKYRME MODEL}
The model starts from employing a map $U$ that maps from $\mathbbm{R}^3$ to the $SU(2)$ group space. Skyrme has determined the form $U$ to be
\begin{equation}
U=\exp(i\xi\hat{n}\cdot\vec{\tau})=\mathbbm{1}.\cos{\xi}+i\left(\hat{n}\cdot\vec{\tau}\right)\sin{\xi}.\label{eq:U}
\end{equation}
The second equality is due to the real-valued unit vector $\hat{n}$ and the properties of Pauli matrices $\vec{\tau}$ (commutation relation $[\tau_a,\tau_b]=2i\epsilon_{abc}\tau_c$ and trace identity $\Tr(\tau_a\tau_b)=2\delta_{ab}$). Here $\xi$ is a real-valued function and $\mathbbm{1}$ is just $2\times2$ identity matrix.

It is usual to use the stereographic projection projected from the north pole of $S^2$ for the unit vector
\begin{equation}
\hat{n}={1\over 1+u\bar{u}}\left(u+\bar{u},-i(u-\bar{u}),1-u\bar{u}\right),
\label{eq:stereographicprojN}
\end{equation}
where $u$ is a complex function and $\bar{u}$ is its complex conjugate.
For compactness, it is a common practice to use the following notation
\begin{equation}
L_\mu=U^\dagger\partial_\mu U,
\end{equation}
called the left-invariant one-form whose value lie in the algebra of $su(2)$. The symmetry property have the known topological degree $B=\int d^3x~\mathcal{B}^0$ from the following definition of baryon current in terms of $\xi,u,$ and $\bar{u}$~\cite{Adam:2015ele}
\begin{equation}
\mathcal{B}^\mu={1\over 24\pi^2}{\epsilon^{\mu\nu\rho\sigma}\over\sqrt{-\det g_{\mu\nu}}}\Tr(L_\nu L_\rho L_\sigma)={(-i)\over \pi^2}{\sin^2\xi\over(1+u\bar{u})^2} {\epsilon^{\mu\nu\rho\sigma}\over\sqrt{-\det g_{\mu\nu}}}
\xi_\nu u_\rho \bar{u}_\sigma.
\end{equation}
The subscript for $\xi$, $u$, and $\bar{u}$ means derivative with respect to the coordinates, e.g. $\xi_\mu\equiv\partial\xi/\partial x^\mu$. 
Throughout this paper, we use metric signature $(+,-,-,-)$.
The generalized Skyrme model has the following Lagrangian density
\begin{equation}
\mathcal{L}=\mathcal{L}_2+\mathcal{L}_4+\mathcal{L}_6+\mathcal{L}_0,\label{eq:LagrFull}
\end{equation}
where
\begin{eqnarray}
\mathcal{L}_2&=&-{1\over2}g^{\mu\nu}~\Tr(L_\mu L_\nu)=\xi_\mu\xi^\mu+{4\sin^2\xi\over(1+u\bar{u})}u_\mu\bar{u}^\mu,
\\
\mathcal{L}_4&=&{1\over16}g^{\mu\rho}g^{\nu\sigma}~\Tr([L_\mu,L_\nu][L_\rho,L_\sigma])
\nonumber\\&=&
-{4\sin^2\xi\over (1+u\bar{u})^2}\left(
\xi_\mu\xi^\mu u_\nu\bar{u}^\nu - \xi_\mu\xi^\nu u_\nu\bar{u}^\mu
+{\sin^2\xi\over(1+u\bar{u})^2}
((u_\mu\bar{u}^\mu)^2-u_\mu u^\mu \bar{u}_\nu\bar{u}^\nu)\right),
\\
\mathcal{L}_6&=&-\lambda^2\pi^4g_{\mu\nu}\mathcal{B}^\mu\mathcal{B}^\nu
\nonumber\\&=&
\lambda^2 {\sin^4\xi\over(1+u\bar{u})^4}g_{\mu\nu}
\left({\epsilon^{\mu\kappa\rho\sigma}\over\sqrt{-\det g_{\mu\nu}}}\xi_\kappa u_\rho \bar{u}_\sigma\right)
\left({\epsilon^{\nu\alpha\beta\gamma}\over\sqrt{-\det g_{\mu\nu}}}
\xi_\alpha u_\beta \bar{u}_\gamma\right),
\\
\mathcal{L}_0&=&-\mu^2 V(\Tr(U))=-\mu^2 V(\cos\xi).
\end{eqnarray}
Here we omit the coupling constants $f_\pi$ and $e$ in which they can be reintroduced by rescaling the length and energy units. We do not specify the potential $V$ and in general we assume $V\equiv V(\xi)$ that will be determined later by the BPS Lagrangian method. In this article, we shall use the following ansatz, in spherical coordinates,
\begin{equation}
\xi=\xi(r),\qquad u(\theta,\varphi)=f(\theta)\exp(ig(\varphi)),
\label{eq:ansatz}
\end{equation}
in which the metric is given by $g_{\mu\nu}=\diag(1,-1,-r^2,-r^2\sin^2\theta)$. Using this ansatz, we obtain
\begin{eqnarray}
\mathcal{B}^0&=&-{2\over \pi^2}{f \sin^2\xi \over(1+f^2)^2 r^2\sin\theta} \xi'(r) f'(\theta) g'(\varphi),~~\mathcal{B}^m=0,
\\
\mathcal{L}_2&=&\mathcal{L}^{(1)}_2+\mathcal{L}^{(2)}_2+\mathcal{L}^{(3)}_2,\nonumber\\
&=&-{4\sin^2\xi\over(1+f^2)^2 r^2}{f'(\theta)^2}-{4f^2 \sin^2\xi\over(1+f^2)^2 r^2\sin^2\theta}{g'(\varphi)^2}-\xi'(r)^2,
\\
\mathcal{L}_4&=&\mathcal{L}^{(1)}_4+\mathcal{L}^{(2)}_4+\mathcal{L}^{(3)}_4,\nonumber\\
&=&-{4f^2 \sin^2\xi\over (1+f^2)^2 r^2\sin^2\theta} \xi'(r)^2 g'(\varphi)^2
-{4\sin^2\xi\over (1+f^2)^2 r^2}\xi'(r)^2 f'(\theta)^2\nonumber\\
&&-{16f^2 \sin^4\xi\over(1+f^2)^4 r^4\sin^2\theta}f'(\theta)^2 g'(\varphi)^2,\\
\mathcal{L}_{6}&=&-\lambda^2 {4 f^2\sin^4\xi\over(1+f^2)^4 r^4\sin^2\theta}\xi'(r)^2 f'(\theta)^2 g'(\varphi)^2,
\end{eqnarray}
where we have used an apostrophe which is defined as taking derivative of a function over its argument, e.g. $\xi'(r)\equiv {\partial\xi \over\partial r}, f'(\theta)\equiv {\partial f\over\partial\theta}, g'(\varphi)\equiv{\partial g \over \partial\varphi}$. In most of the cases, boundary conditions are $\xi(r\to0)=\pi$ and $\xi(r\to\infty)=0$ must be satisfied to ensure that $U$ is well defined at the origin, as well required by topology, and reaches the vacuum, $U=\mathbbm{1}$, near the boundary. For a particular ansatz, $f=\tan(\theta/2)$ and $~g=n \varphi$, it can be shown that an integer $n$ is simply the topological degree $B=\int d^3x~ \mathcal{B}^0$.

\section{THE BPS LAGRANGIAN METHOD}

In this section we will review the BPS Lagrangian method in~\cite{Atmaja:2015umo}. Using static energy $E=-\int d^3x~\mathcal{L}$ of any static system, if the Lagrangian density has Bogomolny equations, one obtains
\begin{equation}
\mathcal{L}=(\textnormal{Squared terms})+\mathcal{L}_{BPS},
\end{equation}
where BPS Lagrangian density $\mathcal{L}_{BPS}$ supposedly contains (only) boundary terms\footnote{What we mean by boundary terms are terms that gives trivial Euler-Lagrange equations.}. Suppose the Lagrangian density $\mathcal{L}$ of $N$ fields $\phi_1,...,\phi_N$ has at most square of first derivative of the fields, $\partial \phi_i$ ($i=1,...,N$), as such the squared terms is taking the following form\footnote{In more general situation $f_i$ may also depend on first derivative of other fields $\partial\phi_j$, where $j\neq i$.}
\begin{equation}\label{squared terms}
(\textnormal{Squared terms})\propto \sum_{i=1}^N \left(\partial \phi_i - f_i(\phi_1,...,\phi_N;\vec{x}) \right)^2,
\end{equation}
where $f_i$ may depend explicitly on coordinates $\vec{x}$. In the BPS limit,
\begin{equation}\label{BPS eqns}
\partial \phi_i = f_i(\phi_1,...,\phi_N;\vec{x})
\end{equation}
and $\mathcal{L}-\mathcal{L}_{BPS}\to0$.
These first-order equations are called Bogomolny equations, which satisfy the Euler-Lagrange equations derived from $E$. Following procedure in BPS Lagrangian method~\cite{Atmaja:2015umo}, these Bogomolny equations can be obtained by subtracting the effective Lagrangian density with the BPS Lagrangian density and setting it to zero, $\mathcal{L}-\mathcal{L}_{BPS}= 0$. With the squared terms taking the form of (\ref{squared terms}), we may consider it as a (mutually) quadratic equation of first derivative of the fields, $\partial\phi_i$ with $i=1,\ldots,N$. Solving it and equating the (two) solutions for every $\partial\phi_i$ give us the Bogomolny equations (\ref{BPS eqns}) and some constraint equations. 

Being boundary terms, the BPS Lagrangian density can be recasted into 
\begin{equation}
\int d^3x \sqrt{\det(g_{mn})} ~\mathcal{L}_{BPS}=\int d^3x ~\partial_m J^m ~~ (m,n=1,2,3),
\end{equation}
where we have used the Einstein summation index with $m$ and $n$ are the spatial indices.
With a suitable set of ansatz, we may set $J^m$ to be explicitly independent of the spatial coordinates, and thus $\partial_m J^m=(\partial J^m/ \partial \phi_i)\partial_m\phi_i$.
In many cases, $\sqrt{\det(g_{mn})} ~\mathcal{L}_{BPS}$ turns out to depend effectively on one coordinate, e.g. radial coordinate in the spherical coordinates, while depedency over the remaining coordinates appear as numerical factors. 

Now we extend the recipe a little more. At first we observe that the BPS Lagrangian density, with previously defined $J^m$, can be rewritten as
\begin{equation}
 \mathcal{L}_{BPS}={1\over\sqrt{\det(g_{mn})}} {\partial J^m(\phi_1,\ldots,\phi_N)\over \partial \phi_i}~ \partial_m\phi_i
\end{equation}
which is proportional to first power of $\partial\phi_i$. Since the Lagrangian density $\mathcal{L}$ contains also square of $\partial\phi_i$, we could ask a question if there are other possible boundary terms that proportional to higher power of $\partial\phi_i$. These possible boundary terms has been studied in~\cite{Adam:2016ipc} and as an example in our case, with $N=3$, it is given by
\begin{equation}
 \mathcal{L}_{BPS}={1\over\sqrt{\det(g_{mn})}} J_{[ijk]}^{[lmn]}(\phi_1,\phi_2,\phi_3)~ \partial_l\phi_i \partial_m\phi_j \partial_n\phi_k,
\end{equation}
where indices inside $[\cdots]$ are totally antisymmetric. In general we can write the BPS Lagrangian density as polynomial function of $\partial\phi_i$ in which the ``constants'' are functions of $\phi_i$, or may be also explicitly of coordinates $\vec{x}$. The boundary terms are defined as the BPS Lagrangian density that gives trivial the Euler-Lagrange equations. As an example for the BPS Lagrangian density with only first power of $\partial\phi_i$, we can write
\begin{equation}
 \mathcal{L}_{BPS}={1\over\sqrt{\det(g_{mn})}} Q^m_i(\phi_1,\ldots,\phi_N)~ \partial_m\phi_i.
\end{equation}
By imposing that the BPS Lagrangian density should produce trivial Euler-Lagrange equations, one can simply show that $Q^m_i={\partial J^m\over \partial \phi_i}$ with $J^m\equiv J^m(\phi_1,\ldots,\phi_N)$. This generalization rises a question if we are allowed to add non-boundary terms, that produce non-trivial Euler-Lagrange equations, into the BPS Lagrangian density. Adding these terms into the BPS Lagrangian density will produce more constraint equations which are basically the Euler-Lagrange equations of these terms. An interesting feature of these non-boundary terms is that the resulting Bogomolny equations can be shown to produce non-zero stress tensor, for more details see~\cite{Atmaja:2018ddi}.

\subsection{BPS Lagrangian density with boundary terms}

We observe that the Lagrangian density (\ref{eq:LagrFull}), in spherical coordinates with ansatz (\ref{eq:ansatz}), contains at most squared of $\xi'(r)$, $f'(\theta)$, and $g'(\varphi)$. Notice that the BPS Lagrangian density with boundary terms, as disscussed in the previous section, contains at most first power of $\xi'(r)$, $f'(\theta)$, and $g'(\varphi)$. Thus general form of the BPS Lagrangian density with boundary terms at least contain the following terms 
\begin{eqnarray}
\mathcal{L}_{BPS}&\propto& Q_\xi~\xi'(r) + Q_f~f'(\theta) + Q_g~g'(\varphi) + Q_{\xi f}~\xi'(r) f'(\theta) + Q_{\xi g}~\xi'(r) g'(\varphi) + Q_{fg}~f'(\theta) g'(\varphi) \nonumber\\
&&+ Q_{\xi fg}~\xi'(r) f'(\theta) g'(\varphi),\label{eq:notenoughQ}
\end{eqnarray}
where $Q_X$ is a function of all the effective fields in $X\equiv X(\xi,f,g)$, which does not depend explicitly on the coordinates; e.g. $Q_\xi$ is a function only of $\xi$ while $Q_{\xi f g}$ is a function of $\xi,f,$ and $g$. However, this form of BPS Lagrangian density is not sufficient in deriving Bogomolny equations for each submodels of the generalized Skyrme model as we will show it in the cases of known submodels~\cite{Adam:2010fg,Harland:2013rxa,Stepien:2015naq,Stepien:2018mti}

\subsection{Setup of BPS Lagrangian density for The Generalized Skyme model}
At the end we expect to have at most three independent BPS equations since there are three first-derivative of effective fields: $\xi'(r)$, $f'(\theta)$, and $g'(\varphi)$. Therefore the submodels must consist of at most three terms containing first-derivative of effective fields with additional potential term, $\mathcal{L}_0$, that can always be included in the calculation. To construct the BPS Lagrangian density we consider each terms in the submodels as independent first-derivative of effective fields labeled by $X,Y,Z$ which are functions of $\xi'(r)$, $f'(\theta)$, and $g'(\varphi)$. Therefore, the effective Lagrangian density that we would consider is generaly given by
\begin{equation}
 \mathcal{L}_{eff}= Cx ~X^2+ Cy~ Y^2+Cz~ Z^2+\mathcal{L}_0,
\end{equation}
where $Cx,Cy,Cz$ are functions of the effective fields $(\xi,f,g)$ and also the coordinates explicitly. The BPS Lagrangian density is then given by
\begin{equation}\label{gen BPS Lagrangian}
 \mathcal{L}_{BPS}={-1\over r^2\sin\theta} \left(Q_0+ Q_x~X+ Q_y~Y+ Q_z~Z+ Q_{xy}~X~Y+ Q_{xz}~X~Z+ Q_{yz}~Y~Z\right),
\end{equation}
where all the $Q$'s are in general functions of the all effective fields, but not explicitly of the coordinates. To be more explicit we will show in more detail how to construct the BPS Lagrangian density of some known submodels of the generalized Skyrme model.

\section{SOME KNOWN BPS SUBMODELS}
The BPS submodels, the genuine and non-genuine ones, of the generalized Skyrme model have been studied in the literature by various methods. In this section, we would like to study them using the BPS Lagrangian method following the setup that we described previously. We will devide them into three groups based on the (effective) number of derivative-terms, from one to three derivative-terms, and show that the non-vacuum solutions found in the literature are indeed solutions to the resulting Bogomolny's equations along with possible constraint equations.

\subsection{One Derivative-term: The BPS Skyrme model}
The Lagrangian density has the following form
\begin{equation}
\mathcal{L}_{eff}=\mathcal{L}_6+\mathcal{L}_0
=-\lambda^2 {4 f^2\sin^4\xi\over(1+f^2)^4 r^4\sin^2\theta}X^2-\mu^2 V(\xi),
\end{equation}
with $X\equiv \xi'(r) f'(\theta) g'(\varphi)$ In this case the BPS Lagrangian density is
\begin{equation}
\mathcal{L}_{BPS}=- {Q_x\over r^2 \sin\theta} X.
\end{equation}
 In the case at hand we omit the $Q_0$ term to simplify calculations in reproducing results in all the known submodels. So solutions of $\mathcal{L}_{eff}-\mathcal{L}_{BPS}=0$ for $X$ are
\begin{equation}
X_\pm = \frac{\csc ^4(\xi )}{8 f^2 \lambda ^2}\left(f^2+1\right)^2 r^2 \sin(\theta )  \left(\left(f^2+1\right)^2 Q_x \pm\sqrt{D}\right),
\end{equation}
with
\begin{equation}
D=f^8 Q_x^2+4 f^6 Q_x^2+6 f^4 Q_x^2+4 f^2 Q_x^2+8 f^2 \lambda ^2 \mu ^2 V \cos (2 \xi )-2 f^2 \lambda ^2 \mu ^2 V \cos (4 \xi )-6 f^2 \lambda ^2 \mu ^2 V+Q_x^2.
\end{equation}
The solutions will be equal if $D=0$, which then gives us 
\begin{equation}
Q_x=\pm \frac{4 f \lambda  \mu  \sqrt{V} \sin ^2(\xi )}{\left(f^2+1\right)^2}.
\end{equation}
So we obtain the Bogomolny's equation
\begin{equation}\label{BEqn known one derivative}
\xi'(r) f'(\theta) g'(\varphi)=\pm \frac{\left(f^2+1\right)^2 \mu  r^2 \sqrt{V} \sin (\theta ) \csc ^2(\xi )}{2 f \lambda }.
\end{equation}
This is in agreement with the result in~\cite{Adam:2010fg} when we substitute $g=n\varphi$, $f=\tan(\theta/2)$, and $V=1-\cos(\xi)$, with $n$ is an integer constant.

\subsection{Two Derivative-Terms: The First BPS Submodel}\label{The First BPS submodel}
There are two BPS submodels that have been identified to have Bogomolny equations, with non-vacuum solutions, in the Skyrme model. In this section we discuss first the First BPS submodel with the following Lagrangian density~\cite{Adam:2017pdh}
\begin{equation}
 \mathcal{L}_{eff}=\mathcal{L}^{(1)}_2+\mathcal{L}^{(2)}_2+\mathcal{L}^{(1)}_4+\mathcal{L}^{(2)}_4=-{4\sin^2\xi\over(1+f^2)^2 r^2}X^2-{4f^2 \sin^2\xi\over(1+f^2)^2 r^2\sin^2\theta}Y^2,
\end{equation}
with $X\equiv\left( f'(\theta)\sqrt{1+\xi'(r)^2}\right)$ and $Y\equiv\left( g'(\varphi)\sqrt{1+\xi'(r)^2}\right)$. Although the $\mathcal{L}_{eff}$ consists of four derivative-terms, it can be simplified into two derivative-terms as shown above. With this $\mathcal{L}_{eff}$, the BPS Lagrangian density is given by
\begin{equation}
 \mathcal{L}_{BPS}={-1\over r^2\sin\theta} \left(Q_x~X+ Q_y~Y+ Q_{xy}~X~Y\right).
\end{equation}
Solving $\mathcal{L}_{eff}-\mathcal{L}_{BPS}=0$ as quadratic equation of $X$ gives solutions
\begin{equation}
 X_\pm=\frac{1}{8}\csc (\theta )\csc ^2(\xi )\left(\left(f^2+1\right)^2 (Y Q_{xy}+Q_x)\pm\sqrt{D_1}\right),
\end{equation}
with 
\begin{equation}\label{D1}
 D_1=\left(f^2+1\right)^4 (Y Q_{xy}+Q_x)^2+16~ Y \sin ^2(\xi ) \left(\left(f^2+1\right)^2 Y \sin (\theta )-4 f^2 \sin ^2(\xi )Q_y\right)
\end{equation}
 Two solutions are equal if $D_1=0$ and it is considered as quadratic equation of $Y$ with solutions
\begin{equation}
 Y_\pm=-\frac{\left(f^2+1\right)^4 Q_x^2}{\left(f^2+1\right)^4 Q_{xy} Q_x+8 \left(f^2+1\right)^2 Q_y \sin (\theta ) \sin ^2(\xi )\pm 4 \sqrt{D_2}}
\end{equation}
with
\begin{equation}
 D_2=\left(f^2+1\right)^6 Q_{xy} Q_x Q_y \sin (\theta ) \sin ^2(\xi )+4 \left(f^2+1\right)^4 \sin ^4(\xi ) \left(f^2 Q_x^2+Q_y^2 \sin ^2(\theta )\right).
\end{equation}
The two solutions will be equal if $D_2=0$ which implies $Q_x=Q_y=0$. This however implies solutions $Y_\pm=0$, but it is actually superficial. If there is any of the $Q$'s is zero then it is suggested to repeat the BPS Lagrangian method from beginning after eliminating all the zero $Q$'s in the BPS Lagrangian density. This is because our previous assumption about the form of BPS Lagrangian density is invalid and this may lead to non-existence of some Bogomolny's equations. 

Repeating the steps starting from $\mathcal{L}_{eff}-\mathcal{L}_{BPS}=0$, with $Q_x=Q_y=0$, we obtain two solutions
\begin{equation}
 X_\pm=\frac{Y}{8} \csc (\theta ) \csc ^2(\xi ) \left(\left(f^2+1\right)^2 Q_{xy}\pm\sqrt{D_1}\right),
\end{equation}
with $D_1=-8 f^2 (\cos (4 \xi )-4 \cos (2 \xi ))+\left(f^2+1\right)^4 Q_{xy}^2-24 f^2$. Again two solutions will be equal if $D_1=0$ and thus gives solutions
\begin{equation}
 Q_{xy}=\pm \frac{8 f}{\left(f^2+1\right)^2 \csc ^2(\xi )}.
\end{equation}
So now the Bogomolnyi's equation is
\begin{equation}
\qquad f'(\theta)=\pm g'(\varphi) f \csc(\theta ).
\end{equation}
Nontrivial solutions to this Bogomolny's equation are $g=n\varphi$ and $f=\tan\left(\theta\over 2\right)^n$, with $n=\pm1,\pm2,\pm3,\dots$.
There is additional constraint equation since Euler-Lagrange equation of $\mathcal{L}_{BPS}$, for $\xi$, is not trivial. Writing 
\begin{equation}
 \mathcal{L}_{BPS}=\pm\frac{8 f}{\left(f^2+1\right)^2 \csc ^2(\xi )r^2\sin(\theta)}f'(\theta) g'(\varphi) \left(1+\xi'(r)^2\right),
\end{equation}
the constraint equation pulled out from its Euler-Lagrange equation for $\xi$ is
\begin{equation}\label{BEqn First BPS submodel}
 \xi ''(r) \sin (\xi (r))+\left(\xi '(r)^2-1\right) \cos (\xi (r))=0.
\end{equation}
One of the solutions to this constraint equation is given in~\cite{Adam:2017pdh} as a compacton. Another possible solution, that satisfies the boundary conditions, is
\begin{equation}
\xi=\begin{cases}
\cos ^{-1}\left(\sin(r)-\cos(r)\right),\qquad 0\leq r \leq {\pi\over 2}\\
0,\qquad\qquad\qquad\qquad\qquad\qquad r> {\pi\over 2}
\end{cases},
\end{equation}
which is also a compacton with radius size is half than the compacton in~\cite{Adam:2017pdh}.

This First BPS submodel can actually be decomposed into two BPS subsubmodels derived using the concept of strong necessary condition~\cite{Stepien:2015naq}. Here we will show that Bogomolny equations of these BPS subsubmodels can also be derived using the BPS Lagrangian method.

\subsubsection{The First BPS subsubmodel}\label{1subsubmodel}
The effective Lagrangian density has the form
\begin{equation}
\mathcal{L}_{eff}=\mathcal{L}^{(1)}_2+\mathcal{L}^{(1)}_4=-\frac{4 \sin ^2(\xi ) }{r^2\left(1+f^2\right)^2}\left(X^2+\frac{f^2}{\sin ^2(\theta )}Y^2\right),
\end{equation}
with $X\equiv f'(\theta)$ and $Y\equiv g'(\varphi) \xi'(r)$. Thus the corresponding BPS Lagrangian density is given by
\begin{equation}
\mathcal{L}_{BPS}=-\frac{1}{r^2 \sin (\theta )}\left(Q_x X+Q_y Y +Q_{xy}X~Y\right).
\end{equation}
To find the Bogomolny equations, we first consider $\mathcal{L}_{eff}-\mathcal{L}_\text{BPS}=0$ as quadratic equation of $X$ in which its solutions are
\begin{eqnarray}
X_\pm= \frac{1}{8} \csc (\theta ) \csc ^2(\xi ) \left\{\left(f^2+1\right)^2 \left(Q_{xy} Y+Q_x\right)\pm \sqrt{D_1}\right\},
\end{eqnarray}
with
\begin{equation}
 D_1=\left(f^2+1\right)^4 ( Q_{xy}Y+Q_x)^2+16~ \sin ^2(\xi ) \left\{2 f^2 Y^2 (\cos (2 \xi )-1)+\left(f^2+1\right)^2 Q_y Y \sin (\theta )\right\}
\end{equation}
must be zero for two solutions to be equal. Taking $D_1=0$ as a quadratic equation of $Y$, we find two solutions
\begin{equation}\label{noexist}
Y_\pm= \frac{-\left(f^2+1\right)^4 Q_x Q_{xy}-8 \left(f^2+1\right)^2 Q_y \sin (\theta ) \sin ^2(\xi )\pm 4 \sqrt{\left(f^2+1\right)^2 \sin ^2(\xi ) D_2}}{\left(f^2+1\right)^4 Q_{xy}^2-64 f^2 \sin ^4(\xi )},
\end{equation}
with
\begin{equation}
 D_2=4 f^2\left(f^2+1\right)^2 Q_x^2 \sin^2 (\xi )+\left(f^2+1\right)^4 Q_{xy} Q_x Q_y \sin (\theta )+4\left(f^2+1\right)^2 Q_y^2 \sin ^2(\theta ) \sin ^2(\xi )
\end{equation}
which again must be zero for two solutions to be equal. We must find all $Q$'s that give $D_2=0$ everywhere; for any values of $r$, $\theta$, and $\varphi$. Notice that there are terms in $D_2$ that depend on spatial coordinates explicitly, and so a natural way to find these $Q$'s are by taking each terms in $D_2$ to be zero everywhere. Hence we find the only non-trivial solution is if $Q_x=Q_y=0$.

After eliminating terms containing $Q_x$ and $Q_y$ in the BPS Lagrangian density, we set again $\mathcal{L}_{eff}-\mathcal{L}_\text{BPS}=0$. From it, we find two solutions for $Y$ as follows
\begin{equation}
Y_\pm= {\csc ^2(\xi )\over 8 f^2} f'(\theta) \sin (\theta )\left(\left(f^2+1\right)^2 Q_{xy}\pm\sqrt{D_1}\right),
\end{equation}
where 
\begin{equation}
D_1=\left(f^8 Q_{xy}^2+4 f^6 Q_{xy}^2+6 f^4 Q_{xy}^2+32 f^2 \cos (2 \xi )-8 f^2 \cos (4 \xi )+4 f^2 Q_{xy}^2-24 f^2+Q_{xy}^2\right).
\end{equation}
Two solutions will be equal if $D_1=0$ which then give us
\begin{equation}
Q_{xy}=\pm\frac{8 f \sin ^2(\xi )}{\left(f^2+1\right)^2}.\label{eq:Qboundary}
\end{equation}
Substituting this into the solutions of $Y$, we obtain the Bogomolny's equation
\begin{equation}
\xi'(r) g'(\varphi)=\pm\frac{f'(\theta) \sin (\theta )}{f}.
\label{eq:BPS1subsub}
\end{equation}
It is easy to prove that the BPS Lagrangian density is a boundary term and so the Bogomolny's equation satisfies the Euler-Lagrange equations.

\subsubsection{The Second BPS subsubmodel}\label{2subsubmodel}
The effective Lagrangian density is given by
\begin{equation}
\mathcal{L}_{eff}=\mathcal{L}^{(2)}_2+\mathcal{L}^{(2)}_4=-\frac{4 \sin ^2(\xi ) }{\left(f^2+1\right)^2}\left(\frac{f^2 }{r^2 \sin ^2(\theta )}X^2+\frac{Y^2}{r^2}\right), 
\end{equation}
with $X\equiv g'(\varphi)$ and $Y\equiv\left( \xi'(r) f'(\theta)\right) $ 
The corresponding BPS Lagrangian density for this subsubmodel is
\begin{equation}
\mathcal{L}_{BPS}=-\frac{Q_x}{r^2 \sin (\theta )}X-\frac{Q_y}{r^2 \sin (\theta )}Y-\frac{Q_{xy}}{r^2 \sin (\theta )}X~Y.
\end{equation}
Solving $\mathcal{L}_{eff}-\mathcal{L}_{BPS}=0$, two solutions for $X$ are
\begin{equation}
X_\pm=\frac{\left(f^2+1\right)^2 \sin (\theta ) (Q_x+ Q_{xy}Y)\pm\sqrt{\sin (\theta ) D_1}}{8 f^2},
\end{equation}
where
\begin{equation}
D_1=\left(f^2+1\right)^4 \sin (\theta ) (Q_x+Q_{xy}Y)^2+16 \sin ^2(\xi ) \left(\left(f^3+f\right)^2 Q_y-4 f^2 \sin (\theta ) \sin ^2(\xi )Y\right)Y.
\end{equation}
Setting $D_1=0$, we obtain two solutions for $Y$,
\begin{equation}
Y_\pm= \frac{-\left(f^2+1\right)^4 Q_x Q_{xy}-4 \csc (\theta ) \left(2 f^2 \left(f^2+1\right)^2 Q_y \sin ^2(\xi )\pm f \left(f^2+1\right)^2 \sin(\xi )\sqrt{D_2}\right)}{\left(f^2+1\right)^4 Q_{xy}^2-64 f^2 \sin ^4(\xi )},
\end{equation}
with
\begin{equation}
D_2=\left(\left(f^2+1\right)^2 Q_x Q_y Q_{xy} \sin (\theta )+4 f^2 Q_y^2 \sin ^2(\xi )+4 Q_x^2 \sin ^2(\theta ) \sin ^2(\xi )\right).
\end{equation}
Similary as in the case of previous subsubmodel, the only non-trivial solution for $D_2=0$ is if $Q_x=Q_y=0$.

Again after removing the zero $Q$'s in the BPS Lagrangian density and setting again $\mathcal{L}_{eff}-\mathcal{L}_{BPS}=0$, we find two solutions for $X$ as follows
\begin{equation}
X_\pm=\frac{1}{8 f^2}\csc ^2(\xi ) \sin (\theta )Y\left(\left(f^2+1\right)^2 Q_{xy}\pm\sqrt{D_1}\right),
\end{equation}
with
\begin{equation}
D_1=\left(f^8 Q_{xy}^2+4 f^6 Q_{xy}^2+6 f^4 Q_{xy}^2+32 f^2 \cos (2 \xi )-8 f^2 \cos (4 \xi )+4 f^2 Q_{xy}^2-24 f^2+Q_{xy}^2\right).
\end{equation}
With only $Q_{xy}$ in $D_1$, the solution to $D_1=0$ is equal to \eqref{eq:Qboundary}. This then imply the Bogomolny equation
\begin{equation}
g'(\varphi)=\pm{\xi'(r) f'(\theta) \sin\theta\over f}.
\label{eq:BPS2subsub}
\end{equation}
As in the previous subsubmodel, the BPS Lagrangian density is a boundary term and so this Bogomolny equation satisfies the Euler-Lagrange equations.

The Bogomolny's equations of (\ref{eq:BPS1subsub}) and (\ref{eq:BPS2subsub}) imply that solutions for $g(\varphi)$ and $\xi(r)$ are linear functions of $\varphi$ and $r$, respectively. With the right boundary condition for $f$,
\begin{equation}
 f(\theta =0)=0,\qquad f(\theta=\pi)\to\infty,
\end{equation}
solutions for $f$ are
\begin{equation}
 f(\theta)\propto \left(\tan\left(\theta\over 2\right)\right)^{c_1},
\end{equation}
where $c_1>0$ is a constant related to constant slopes in the linear solutions of $g(\varphi)$ and $\xi(r)$. Explicit solutions were given in~\cite{Stepien:2018mti} and they are known as compacton~\cite{Adam:2017pdh}.

\subsection{Two Derivative-Terms: The Second BPS Submodel}
Although the second BPS submodel in~\cite{Adam:2017pdh} was not derived using the concept of strong necessary condition, here we would like to show that its Bogomolny equation can also be derived using the BPS Lagrangian method. 
The effective Lagrangian density of the second BPS submodel is
\begin{equation}
\mathcal{L}_{eff}=\mathcal{L}^{(3)}_2+\mathcal{L}^{(3)}_4=-X^2-{16f^2 \sin^4\xi\over(1+f^2)^4 r^4\sin^2\theta}Y^2,
\end{equation}
with $X\equiv\xi'(r)$ and $Y\equiv f'(\theta) g'(\varphi)$. So the corresponding BPS Lagrangian density is
\begin{equation}
\mathcal{L}_{BPS}=-{Q_x\over r^2\sin\theta}X-{Q_y\over r^2 \sin\theta}Y-{Q_{xy}\over r^2 \sin\theta}X~Y.
\end{equation}
Equating both, we obtain two solutions for $X$,
\begin{equation}
X_\pm=\frac{\csc (\theta )}{2 \left(f^2+1\right)^2 r^2} \left(\left(f^2+1\right)^2 (Q_x+ Q_{xy} Y)\pm\csc (\theta )\sin(\theta ) \sqrt{D_1}\right)
\end{equation}
with
\begin{eqnarray}
D_1&=&\left(f^2+1\right)^4 Y^2 Q_{xy}^2+Q_x^2\nonumber\\
&&+4 Y \left(\left(f^2+1\right)^4 Q_y r^2 \sin (\theta )-2 f^2 Y (\cos (4 \xi )-4 \cos (2 \xi ))\right)\nonumber\\
&&+2 \left(f^2+1\right)^4 Y Q_x Q_{xy}+f^2 \left(\left(f^6+4 f^4+6 f^2+4\right) Q_x^2-24 Y^2\right).
\end{eqnarray}
From setting $D_1=0$, in order for the two solutions to be equal, we then obtain two solutions for $Y$,
\begin{equation}
Y_\pm=\frac{-2 \left(f^2+1\right)^4 Q_y r^2 \sin (\theta )-\left(f^2+1\right)^4 Q_x Q_{xy}\pm 2 \sqrt{D_2}}{-8 f^2 (\cos (4 \xi )-4 \cos (2 \xi ))+\left(f^2+1\right)^4 Q_{xy}^2-24 f^2},
\end{equation}
with
\begin{equation}
D_2=\left(f^2+1\right)^8 Q_y^2 r \sin ^2(\theta )+\left(f^2+1\right)^8 Q_y Q_x Q_{xy} r^2 \sin (\theta )+16 f^2 \left(f^2+1\right)^4 Q_x^2 \sin ^4(\xi ).
\end{equation}
Then we set $D_2=0$ to make the two solutions to be equal. We have to solve $D_2=0$ for whole value of spherical coordinates. Expanding it first as a series with respect to $r$ and then solving it for each term in the series. The terms with $r^0$ and $r^1$ give $Q_x=0$ and $Q_y=0$, respectively, leaving $Q_{xy}$ undetermined, which are enough to solve it.

Repeating again the BPS Lagrangian method, with only $Q_{xy}$ is nonzero in the BPS Lagrangian density, then we get two solutions for $X$,
\begin{equation}
X_\pm=\frac{\csc (\theta )}{2 r^2 \left(f^2+1\right)^2 }Y \left(Q_{xy} \left(f^2+1\right)^2 \pm  \csc (\theta )\sin(\theta ) \sqrt{D_1} \right),
\end{equation}
with
\begin{equation}
D_1=\left(-8 f^2 (\cos (4 \xi )-4 \cos (2 \xi ))+\left(f^2+1\right)^4 Q_{xy}^2-24 f^2\right).
\end{equation}
This two solutions will be equal if $D_1=0$ which is solved by the same $Q_{xy}$ given in (\ref{eq:Qboundary}). Upon substituting (\ref{eq:Qboundary}), we obtain the Bogomolny's equation
\begin{equation}
\xi'(r)=\pm \frac{4 f \csc (\theta ) \sin ^2(\xi)}{r^2 \left(f^2+1\right)^2}f'(\theta) g'(\varphi).
\end{equation}
We also check that this Bogomolny equation indeed satisfies the Euler-Lagrange equations and the solutions have been disscused in~\cite{Adam:2017pdh}.

\subsection{Three Derivative-terms: The Skyrme-term with potential}\label{known three-derivative}
The Lagrangian of this model is given by~\cite{Harland:2013rxa}
\begin{equation}
 \mathcal{L}_{eff}=\mathcal{L}_4 +\mathcal{L}_0=-{16f^2 \sin^4\xi\over(1+f^2)^4 r^4\sin^2\theta} X^2-{4\sin^2\xi\over (1+f^2)^2 r^2}Y^2-{4f^2 \sin^2\xi\over (1+f^2)^2 r^2\sin^2\theta}Z^2-V(\xi),
\end{equation}
with $X\equiv\left(f'(\theta) g'(\varphi)\right)$, $Y\equiv\left(f'(\theta)\xi'(r) \right)$, and $Z\equiv\left(g'(\varphi)\xi'(r) \right)$.
Here we have set the coupling $\mu\to1$. The BPS Lagrangian density for this model is given by
\begin{equation}
 \mathcal{L}_{BPS}= \frac{1}{r^2 \sin (\theta )}\left( Q_x X+ Q_y Y+ Q_z Z+Q_{xy} X~Y+ Q_{xz} X~Z+ Q_{yz} Y~Z\right).
\end{equation}
Following the previous steps we solve the equation $\mathcal{L}_{eff}-\mathcal{L}_{BPS}=0$ for $X$, $Y$, and $Z$ subsequently. At the end we obtain a consraint equation which can be written in terms of explicit power of $r$. We solve it by setting each of the terms to zero, which give us three constraint equations: for the $r^{-2}$-term,
\begin{eqnarray}\label{r-2}
 16 \left(f^2+1\right)^4 Q_{yz} Q_y Q_z \csc (\theta ) \csc ^4(\xi )-64 \left(f^2+1\right)^2 \csc ^2(\xi ) \left(f^2 Q_y^2 \csc ^2(\theta )+ Q_z^2\right)=0;\nonumber\\
\end{eqnarray}
for the $r^0$-term,
\begin{eqnarray}\label{r0}
&&-\left(f^2+1\right)^8 \csc ^8(\xi ) \left(-2 Q_x Q_{yz} (Q_{xy} Q_z+ Q_{xz} Q_y)+( Q_{xz} Q_{Y}-Q_{xy} Q_z)^2+Q_x^2 Q_{yz}^2\right)\nonumber\\
&&+8 \left(f^2+1\right)^6 Q_x \csc (\theta ) \csc ^6(\xi ) \left(-2 f^2 Q_{xy} Q_{Y}+Q_{xz} Q_z \cos (2 \theta )- Q_{xz} Q_z\right)\nonumber\\
&&+64 f^2 \left(f^2+1\right)^4 \csc ^4(\xi ) \left(Q_x^2+ Q_{yz}^2 V\right)-4096 f^4 V=0;
\end{eqnarray}
and for the $r^2$-term,
\begin{eqnarray}\label{r2}
 V \left(Q_{xy}^2 \csc ^2(\theta )-\frac{\left(\left(f^2+1\right)^2 Q_{xy} Q_{yz} \csc (\theta ) \csc ^2(\xi )-8 Q_{xz}\right)^2}{\left(f^2+1\right)^4 Q_{yz}^2 \csc ^4(\xi )-64 f^2}\right)=0.
\end{eqnarray}
The constraint equation (\ref{r2}) has solutions $V=0$ or 
\begin{equation}\label{sol Qfgxixi}
 Q_{yz}=\frac{4 \sin (\theta ) \sin ^2(\xi ) \left(f^2 Q_{xy}^2 \csc ^2(\theta )+Q_{xz}^2\right)}{\left(f^2+1\right)^2 Q_{xy} Q_{xz}}.
\end{equation}

Choosing $V=0$, the remaining constraint equations (\ref{r-2}) and (\ref{r0}) imply $Q_y,Q_z,Q_x=0$. Redo the calculation starting from $\mathcal{L}_{eff}-\mathcal{L}_{BPS}=0$, with $V,Q_y,Q_z,Q_x=0$, gives us a constraint equation
\begin{eqnarray}
 &&\csc ^2(\theta ) \left(\left(f^2+1\right)^4 Q_{yz}^2 \csc ^4(\xi )-64 f^2\right) \left(\left(f^2+1\right)^6 Q_{xy}^2 r^2 \csc ^6(\xi )-256 f^2\right)\nonumber\\
 &&-\left(f^2+1\right)^6 \csc ^6(\xi ) \left(\left(f^2+1\right)^2 Q_{xy} Q_{yz} r \csc (\theta ) \csc ^2(\xi )-8 Q_{xz} r\right)^2=0.
\end{eqnarray}
Similarly writing this constraint equation in terms of explicit power of $r$ yields two constraint equations: the $r^0$-term,
\begin{eqnarray}\label{V0r0}
 \left(f^2+1\right)^4 Q_{yz}^2 \csc ^4(\xi )-64 f^2=0;
\end{eqnarray}
and for $r^2$-term,
\begin{equation}\label{V0r2}
 16 \left(f^2+1\right)^2 Q_{xy} Q_{xz} Q_{yz} \csc (\theta ) \csc ^2(\xi )-64 \left(f^2 Q_{xy}^2 \csc ^2(\theta )+Q_{xz}^2\right)=0.
\end{equation}
Unfortunately the constraint equation (\ref{V0r0}) is not allowed since it will imply the Bogomolny's equation for $Z$,
\begin{equation}
 Z=X\csc ^2(\xi )\left(f^2+1\right)^2\frac{8  Q_{xz} \sin (\theta )-\left(f^2+1\right)^2 Q_{xy} Q_{yz} \csc ^2(\xi )}{\left(f^2+1\right)^4 Q_{yz}^2 \csc ^4(\xi )-64 f^2}
\end{equation}
to becomes indeterminate. 

Therefore we prefer to take (\ref{sol Qfgxixi}) as solution for the constraint equation (\ref{r2}). The remaining constraint equations (\ref{r-2}) and (\ref{r0}) imply $Q_z,Q_y=0$ and  
\begin{equation}\label{EL CE}
 64 f^2 V-\left(f^2+1\right)^4 Q_x^2 \csc ^4(\xi )=0.
\end{equation}
In addition there are more constraint equations comming from Euler-Lagrange equations of the corresponding $\mathcal{L}_{BPS}$ since $Q_{xy},Q_{xz},$ and $Q_{yz}$ are still non-zero. It turns out these additional constraint equations require $Q_{xy},Q_{xz}=0$, which would likely imply $Q_{yz}=0$ upon substituting to the equation (\ref{sol Qfgxixi}). However redoing the calculation starting from $\mathcal{L}_{eff}-\mathcal{L}_{BPS}=0$ with $Q_z,Q_y,Q_{xy},Q_{xz}=0$, we will obtain constraint equations (\ref{V0r0}) and (\ref{EL CE}) that fix the functions
\begin{equation}
 Q_x=\pm \frac{8 f  \sqrt{V}}{\left(f^2+1\right)^2 \csc ^2(\xi )}\qquad \text{and}\qquad Q_{yz}=\pm \frac{8 f}{\left(f^2+1\right)^2 \csc ^2(\xi )}.
\end{equation}
The Bogomolny's equations are given by, for $X$,
\begin{equation}\label{BEqn known 3}
 f'(\theta) g'(\varphi)=\pm \frac{\left(f^2+1\right)^2 r^2 \sqrt{V} \sin (\theta ) \csc ^2(\xi )}{4 f}
\end{equation}
and, for $Y$,
\begin{equation}\label{BEqn known 3-1}
 f'(\theta) =\pm f g'(\varphi) \csc (\theta ).
\end{equation}
It should be clear from the Bogomolnyi's equation for $X$ that $r^2 \sqrt{V} \csc ^2(\xi )=\text{constant}$ which could be considered as solutions for $\xi$. The Bogomolny's equation for $Y$ implies $g=n\varphi$ and $f=\tan(\theta/2)^n$, with $n=\pm1,\pm2,\pm3,\ldots$.
There is still one additional constraint equation comming from Euler-Lagrange equations of the corresponding $\mathcal{L}_{BPS}$,
\begin{equation}
 \frac{\sin (\xi) V'(\xi)}{4 \sqrt{V(\xi)}}+\cos (\xi) \sqrt{V(\xi)}=\xi ''(r) \sin (\xi)+\xi '(r)^2 \cos (\xi),
\end{equation}
 and in the BPS limit it can be simplified to
\begin{eqnarray}\label{cons V}
 &&r^2 \left(4 \cot (\xi ) V(\xi )-V'(\xi )\right)^3 \left(V'(\xi )+4 \cot (\xi ) V(\xi )\right)\nonumber\\
 &&-16 \left(V(\xi )^{3/2} \left(-5 V'(\xi )^2+4 V(\xi ) \left(V''(\xi )+\cot (\xi ) V'(\xi )\right)+16 \csc ^2(\xi ) V(\xi )^2\right)\right)=0.
\end{eqnarray}
Solving each terms, in the explicit power of $r$, in this constaint equation gives $V=\sin(\xi)^4$, but this will imply $r=\text{constant}$ upon substituting it into the corresponding Bogomolny's equation. Interestingly there is other solution given in~\cite{Harland:2013rxa} in which $V=\left(1-\cos(\xi)\right)^4$, with $g=\varphi$ and $r^2 \sqrt{V} \csc ^2(\xi )=1$, for topological degree $B=1$. The constraint equation (\ref{cons V}), along with the Bogomolny's equation $r^2 \sqrt{V} \csc ^2(\xi )=\text{constant}$, opens up possibilities that there might be other potentials for Skyrmion with topological degree $B=1$ or even Skyrmion with topological degree $B>1$.

\section{NEW SUBMODELS WITH ONE DERIVATIVE-TERM: Submodel $\mathcal{L}^{(3)}_4+\mathcal{L}_0$}\label{one-derivative}

A first new submodel that we consider here consist of only one derivative-term with the following effective Lagrangian density,
\begin{equation}\label{Leff1}
\mathcal{L}_{eff}=-{16f^2 \sin^4\xi\over(1+f^2)^4 r^4\sin^2\theta}X^2-\mu ^2 V,
\end{equation}
with $X\equiv(f'(\theta) g'(\varphi))$. The BPS Lagrangian density then is given by
\begin{equation}
\mathcal{L}_{BPS}=-{Q_0\over r^2 \sin (\theta )}-{Q_x\over r^2 \sin (\theta )}X.
\end{equation}
From here on, in deriving the Bogomolny's equations of the new submodels, we will always include the potential $V$ in the effective Lagrangian density and the $Q_0$ in the BPS Lagrangian density. 
Using $\mathcal{L}_{eff}=\mathcal{L}_{BPS}$, solutions for $X$ are
\begin{equation}
X_\pm=
\frac{\csc ^4(\xi ) \left(\left(f^2+1\right)^4 Q_x r^2 \sin (\theta )\pm\sqrt{\left(f^2+1\right)^4 r^2 \sin (\theta ) D_1}\right)}{32 f^2} 
\label{eq90}
\end{equation}
with
\begin{equation}
D_1= 64 f^2 Q_0 \sin ^4(\xi )+r^2 \sin (\theta ) \left(\left(f^2+1\right)^4 Q_x^2-8 f^2 \mu ^2 V (\cos (4 \xi )-4 \cos (2 \xi ))-24 f^2 \mu ^2 V\right).
\end{equation}
Setting $D_1=0$ and solving it for each terms in explicit power of $r$, we obtain solution $Q_0=0$ and
\begin{equation}
Q_x=\pm\frac{8 f \mu  \sqrt{V} \sin ^2(\xi )}{\left(f^2+1\right)^2}.
\end{equation}
Substituting it back to \eqref{eq90}, we obtain Bogomolny's equation
\begin{equation}\label{BEqn1}
f'(\theta) g'(\varphi)= \pm\frac{\left(f^2+1\right)^2 \mu  r^2 \sqrt{V} \sin (\theta ) \csc ^2(\xi )}{4 f}.
\end{equation}
Since $Q_x$ contains $\xi$, the BPS Lagrangian density may not be a boundary term and thus implies additional constraint equation that would determined the solution for $\xi(r)$. This should not be corect since the effective Lagrangian density (\ref{Leff1}) does not contain first-derivative of $\xi$, $\xi'(r)$, then we should think $\xi$ as a background function. For the case disscused here, the correct background function $\xi(r)$ is determined by the Bogomolny's equation (\ref{BEqn1}) in which $r^2 \sqrt{V} \csc ^2(\xi )=\text{constant}$. Choosing the constant to be $1$, the background function $\xi(r)$ satisfies the correct boundary conditions is given by
\begin{equation}
 \xi(r)=\cos^{-1}\left(r^2-1 \over r^2+1\right),
\end{equation}
with $V=\left(1-\cos(\xi)\right)^4$. With this background function $\xi$, the BPS Lagrangian density becomes a boundary term and so there is no additional constraint equation. One should notice that the Bogomolny's equation (\ref{BEqn1}) and the Bogomolny's equation (\ref{BEqn known 3}) coincide and thus, have the same solutions.

\section{NEW SUBMODELS WITH TWO DERIVATIVE-TERMS}\label{two-derivative}
In this section all submodels will have the same BPS Lagrangian density,
\begin{equation}\label{Lbps two}
\mathcal{L}_{BPS}=-\frac{1}{r^2 \sin (\theta )}\left(Q_0+Q_x X+Q_y Y+ Q_{xy} X~Y\right),
\end{equation}
with definitions for $X$ and $Y$ vary in each submodel and they will be written excplicitly. Following previous steps, we will solve $\mathcal{L}_{eff}-\mathcal{L}_{BPS}=0$ by considering it first as a quadratic equation of $X$ and $Y$ subsequently.

\subsection{Submodel $\mathcal{L}^{(1)}_4+\mathcal{L}^{(2)}_4+\mathcal{L}_0$}\label{first two-derivative}
The effective Lagrangian density is
\begin{equation}
\mathcal{L}_{eff}=-\frac{4 \sin ^2(\xi ) }{\left(f^2+1\right)^2}\left(\frac{X^2}{r^2}+\frac{f^2 Y^2}{r^2 \sin ^2(\theta )}\right)-\mu^2 V,\label{eq:1stsubmodel}
\end{equation}
with $X\equiv\xi'(r) f'(\theta)$ and $Y\equiv\xi'(r)g'(\varphi)$. The Bogomolny's equations are given by
\begin{eqnarray}
X&=&\frac{1}{8} \left(f^2+1\right)^2 \csc (\theta ) \csc ^2(\xi ) (Q_x+Q_{xy} Y),\\
Y&=&\frac{\left(f^2+1\right)^2 \left(\left(f^2+1\right)^2 Q_x Q_{xy} \csc (\theta ) \csc ^2(\xi )+8 Q_y\right)}{\csc (\theta ) \sin ^2(\xi ) \left(64 f^2-\left(f^2+1\right)^4 Q_{xy}^2 \csc ^4(\xi )\right)},
\end{eqnarray}
with a constraint equation
\begin{eqnarray}
\mu ^2 V&=&\frac{\csc (\theta ) \left(\left(f^2+1\right)^2 Q_x^2 \csc (\theta ) \csc ^2(\xi )+16 Q_0\right)}{16 r^2}\nonumber\\
&&-\frac{\left(f^2+1\right)^2 \left(\left(f^2+1\right)^2 Q_x Q_{xy} \csc (\theta ) \csc ^2(\xi )+8 Q_y\right)^2}{16 r^2 \sin ^2(\xi ) \left(\left(f^2+1\right)^4 Q_{xy}^2 \csc ^4(\xi )-64 f^2\right)}.
\end{eqnarray}
The only non-trivial solution for this constraint equation, that is valid in the whole space, is $V=Q_0=Q_x=Q_y=0$. Unfortunately, setting $Q_x=Q_y=0$ will make the Bogomolny's equation for $Y$ to be trivial, $Y=0$. Therefore we need to redo the calculation.

Again redoing the calculation by setting $V=Q_0=Q_x=Q_y=0$ in the effective Lagrangian density and in the BPS Lagrangian density above, we then obtain a Bogomolny's equation,
\begin{equation}
X=\frac{1}{8} \left(f^2+1\right)^2 \csc (\theta ) \csc ^2(\xi ) Q_{xy} Y,
\end{equation}
with
\begin{equation}
Q_{xy}=\pm\frac{8 f \sin ^2(\xi )}{\left(f^2+1\right)^2},\label{eq:Qxxfg}.
\end{equation}
The Bogomolny's equation can be simplified to
\begin{equation}\label{BEqn2}
g'(\varphi) =\pm \frac{f'(\theta) \sin (\theta )}{f}.
\end{equation}
Solutions to this Bogomolny's equation are $g(\varphi)=n\varphi$ and $f(\theta)=\tan\left(\theta\over 2\right)^{n}$, with $n=\pm1,\pm2,\pm3,\ldots$.
In this case, Euler-Lagrange equations of the BPS Lagrangian density are not all trivial. It turns out only the Euler-Lagrange equation for $\xi$ is non-trivial, which then becomes a constraint equation,
\begin{equation}
\xi''(r) \sin (\xi)+\xi'(r)^2 \cos (\xi)=0.\label{eq:xiconstrained}
\end{equation}
One may consider the constraint equation (\ref{eq:xiconstrained}) as Bogomolny's equation for $\xi$, $\xi'(r)\sin(\xi)=\text{constant}$, which has solutions
\begin{equation}
\xi(r)=\pm\cos ^{-1}\left(c_1 (-r)-c_2 c_1\right),
\end{equation}
where $c_1$ and $c_2$ are integration constants. Since $\xi$ is a real valued function, the solutions above are valid within a range of $|\left(c_1 (-r)-c_2 c_1\right)|\leq1$. Considering boundary conditions for $\xi$, possible solutions for $\xi$ are
\begin{equation}\label{compacton}
\xi=\begin{cases}
\cos ^{-1}\left({r\over R}-1\right),\qquad 0\leq r \leq 2R\\
0,\qquad\qquad\qquad\qquad\quad r>2R
\end{cases},
\end{equation}
with $R>0$ is an arbitrary constant related to size of the soliton. These solutions are compacton similar to the one obtained in~\cite{Adam:2017pdh}, but here radius size of the compacton is not fixed and it depends on the constant $R$.

\subsection{Submodel $\mathcal{L}^{(1)}_2+\mathcal{L}^{(2)}_2+\mathcal{L}_0$}
The effective Lagrangian density here contains
\begin{equation}
\mathcal{L}_{eff}=-\frac{4 \sin ^2(\xi ) }{\left(f^2+1\right)^2}\left(\frac{X^2}{r^2}+\frac{f^2 Y^2}{r^2 \sin ^2(\theta )}\right)-\mu^2 V,\label{eq:2ndsubmodel}
\end{equation}
with $X\equiv f'(\theta)$ and $Y\equiv g'(\varphi)$.
One should realize that the effective Lagrangian density, in terms of $X$ and $Y$, is the same as in the case of first submodel in the subsection~\ref{first two-derivative} and hence implies $V=Q_0=Q_x=Q_y=0$ and $Q_{xy}$ is given by (\ref{eq:Qxxfg}). It turns out resulting Bogomolny's equation is the same given by the Bogomolny's equation(\ref{BEqn2}). However here we do not have the constraint equation (\ref{eq:xiconstrained}) since function $\xi(r)$ is considered to be a background function as has been disscused in the section~\ref{one-derivative}. Unlike in the section~\ref{one-derivative}, in which the background function $\xi(r)$ is implicitly constrainted by the Bogomolny's equation (\ref{BEqn1}), here we can choose any background function $\xi(r)$ that satisfies the boundary conditions.

\subsection{Submodel $\mathcal{L}^{(3)}_2+\mathcal{L}_6+\mathcal{L}_0$}\label{third two-derivatives}

Here the effective Lagrangian density is
\begin{equation}
\mathcal{L}=-\lambda^2 {4 f^2\sin^4\xi\over(1+f^2)^4 r^4\sin^2\theta}X^2-Y^2-\mu^2 V,\label{eq:4thsubmodel}
\end{equation}
with $X\equiv\xi'(r) f'(\theta) g'(\varphi)$ and $Y\equiv \xi'(r)$. With this effective Lagrangian desity, the Bogomolny's equations are
\begin{eqnarray}
X&=&\frac{\left(f^2+1\right)^4 r^2 (Q_x+Q_{xy} Y)}{8 f^2 \lambda ^2 \csc (\theta ) \sin ^4(\xi )},\label{BEqn3} \\
Y&=&\frac{\left(f^2+1\right)^4 Q_x Q_{xy} \csc ^4(\xi )}{16 f^2 \lambda ^2-\left(f^2+1\right)^4Q_{xy}^2 \csc ^4(\xi )},\label{BEqn4}
\end{eqnarray}
with a constraint equation
\begin{equation}
 \frac{\left(f^2+1\right)^4 Q_x^2 \csc ^4(\xi )}{16 f^2 \lambda ^2}+\frac{\left(16 f^2 \lambda ^2\right) \left(\frac{\left(f^2+1\right)^4 Q_x Q_{xy} \csc ^4(\xi )}{8 f^2 \lambda ^2}+\frac{Q_y \csc (\theta )}{r^2}\right)^2}{4 \left(16 f^2 \lambda ^2-\left(f^2+1\right)^4 Q_{xy}^2 \csc ^4(\xi )\right)}+\frac{Q_0 \csc (\theta )}{r^2}-\mu ^2 V=0.
\end{equation}
This constraint equation requires $Q_0=Q_y=0$ in which it reduces to
\begin{equation}
 \mu ^2 V=\frac{Q_x^2\left(f^2+1\right)^4 \csc ^4(\xi )}{16 f^2 \lambda ^2-Q_{xy}^2\left(f^2+1\right)^4 \csc ^4(\xi )}.
\end{equation}
Setting $Q_0=Q_y=0$ does not make any of Bogomolny's equations (\ref{BEqn3}) and (\ref{BEqn4}) to be trivial. Hence we do not need to redo the calculation. In this reduced constraint equation, the nonzero value of $Q_x$ is determined by nonzero value of $V$ and vice versa. Substituting
\begin{equation}
 Q_x= \pm \sqrt{{\mu ^2 V\left(16 f^2 \lambda ^2-Q_{xy}^2\left(f^2+1\right)^4 \csc ^4(\xi )\right)\over \left(f^2+1\right)^4 \csc ^4(\xi )} }
\end{equation}
into the Bogomolny's equations yields
\begin{eqnarray}
\xi'(r)&=&\pm Q_{xy}\sqrt{\mu ^2 V \left(f^2+1\right)^4 \csc ^4(\xi )\over \left(16 f^2 \lambda ^2-Q_{xy}^2\left(f^2+1\right)^4 \csc ^4(\xi )\right)},\\
f'(\theta) g'(\varphi)&=&\pm{2r^2\sin(\theta)\over Q_{xy}}.
\end{eqnarray}
From the above Bogomolny's equations, we may conclude $Q_{xy}=\frac{4 f \lambda}{\left(f^2+1\right)^2} Q(\xi)$ and so the Bogomolny's equations become
\begin{eqnarray}
 \xi'(r)&=&\pm Q(\xi) \sqrt{\frac{\mu ^2 V}{\sin ^4(\xi )-Q(\xi)^2}},\\
 f'(\theta) g'(\varphi)&=&\frac{\left(f^2+1\right)^2 r^2 \sin (\theta )}{2 f \lambda  Q(\xi)}.
\end{eqnarray}
Furthermore, these Bogomolny's equations imply $Q(\xi)\propto r^2$. One example solution for $Q(\xi)$ is given by
\begin{equation}
 Q(\xi)=\frac{\sin ^2(\xi )}{(1-2\cos (\xi ))^2},\qquad \xi(r)=\cos^{-1}\left(\frac{2 r^2+\sqrt{c_1 \left(3 r^2+c_1\right)}}{4 r^2+c_1}\right),
\end{equation}
with $c_1>0$ is a real constant. The constant $c_1$ is related to the parameters in solutions of $f(\theta)$ and $g(\varphi)$. If we set $g=n\varphi$ and $f=\tan(\theta/2)$ then $c_1=\pm{n\lambda\over 2}$, with $n$ is an integer constant. There is, however, an additional constraint equation comming from the Euler-Lagrange equation of the BPS Lagrangian density for $\xi$,
\begin{equation}
 \xi '(r)^2 Q'(\xi)+2 Q(\xi) \xi ''(r)=0,
\end{equation}
that will fix the potential to be
\begin{equation}
 V=c_2(1-2 \cos (\xi ))^2 \cos (\xi ) (-2 \cos (\xi )+\cos (2 \xi )+2) \sec ^2\left(\frac{\xi }{2}\right).
\end{equation}
with $c_2>0$ is a real constant. The potential has local minimum at $\xi=\pi/3$ within a range $0\leq \xi \leq \pi/3$. This will change the boundary conditions that we have previously, which could imply the value of topological degree $B$ could be fractional, and so it is out of the scope disscused in this article.

We could try another way to get the Bogomolny's equations by setting $V=0$, or equivalently $Q_x=0$. This way we must redo the calculation with the BPS Lagrangian density
\begin{equation}
\mathcal{L}_{BPS}=-\frac{Q_{xy}}{r^2 \sin (\theta )}\xi^2_r f'(\theta) g'(\varphi).
\end{equation}
We then find the Bogomolny's equation
\begin{equation}
2 f \lambda  \csc (\theta ) \sin ^2(\xi ) f'(\theta) g'(\varphi)=\pm\left(f^2+1\right)^2 r^2
\label{eq:signf2}
\end{equation}
and, from the constraint equation, 
\begin{equation}
Q_{xy}=\pm\frac{4 f \lambda  \sin ^2(\xi )}{\left(f^2+1\right)^2}.
\end{equation}
Additional consraint equatiom from Euler-Lagrange equation of the BPS Lagrangian density for $\xi$ is equal to equation (\ref{eq:xiconstrained}), which then could lead to the compacton solution (\ref{compacton}). However, unlike case in the subsection~\ref{first two-derivative}, the Bogomolny's equation here contains $\xi$ and coordinate $r$ and so implies solution for $\xi$,
\begin{eqnarray}
\xi=\pm \sin ^{-1}\left(\sqrt{c_1} r\right)+ \pi  c_2
\end{eqnarray}
where $c_1>0$ is a real constant and $c_2$ is an integer constant. Unfortunately, this solution satisfies the constraint equation (\ref{eq:xiconstrained}) only if $c_1=0$. Here $\xi$ is a constant and thus identified, with $c_2=0$,  as the vacuum solution of the Skyrme model.

\subsection{Submodel $\mathcal{L}^{(3)}_2+\mathcal{L}^{(3)}_4+\mathcal{L}_6+\mathcal{L}_0$}\label{New BPS submodel II-D}
In this submodel, the effective Lagrangian density is
\begin{equation}
\mathcal{L}=-\lambda^2 {4 f^2\sin^4\xi\over(1+f^2)^4 r^4\sin^2\theta}X^2-Y^2-\mu^2 V,
\end{equation}
which is equal to the effective Lagrangian density (\ref{eq:4thsubmodel}) previously, with $X\equiv\left(\xi'(r) f'(\theta) \sqrt{{4\over\lambda^2}+ g'(\varphi)^2}\right)$ and $Y\equiv \xi'(r)$. We may then use the result of the previous submodel and obtain the nonzero components of the BPS Lagrangian density, for nonzero potential,
\begin{equation}
 Q_x=\pm\frac{4 \lambda  f}{\left(f^2+1\right)^2} \sqrt{\mu ^2 V \left(\sin ^4(\xi )-Q(\xi)^2\right)},\qquad Q_{xy}=\frac{4 \lambda  f~ Q(\xi)}{\left(f^2+1\right)^2}.
\end{equation}
The Bogomolny's equations are simplified to
\begin{eqnarray}
 \xi '(r)&=&\pm Q(\xi ) \sqrt{\frac{\mu ^2 V}{\sin ^4(\xi )-Q(\xi )^2}}\\
 f'(\theta ) g'(\varphi )&=&\pm\frac{\left(f^2+1\right)^2 r^2 \sin (\theta )}{4 f} \sqrt{\frac{4 \mu ^2 V}{4 \left(\sin ^4(\xi )-Q(\xi )^2\right)+\lambda ^2 \mu ^2 V Q(\xi )^2}}.
\end{eqnarray}
There is also an additional constraint equation from the BPS Lagrangian density which, in the BPS limit, is given by
\begin{eqnarray}
 &&Q(\xi )^7 V(\xi ) Q'(\xi ) \left(\lambda ^2 \mu ^2 V(\xi )-4\right)^2-4 \sin ^{12}(\xi ) Q(\xi ) V(\xi ) Q'(\xi ) \left(\lambda ^2 \mu ^2 V(\xi )+4\right)\nonumber\\
 &-&12 \sin ^8(\xi ) Q(\xi )^3 V(\xi ) Q'(\xi ) \left(\lambda ^2 \mu ^2 V(\xi )-4\right)-3 \sin ^4(\xi ) Q(\xi )^5 V(\xi ) Q'(\xi ) \left(\lambda ^2 \mu ^2 V(\xi )-4\right)^2\nonumber\\
 &+&Q(\xi )^8 V'(\xi ) \left(\lambda ^2 \mu ^2 V(\xi ) \left(\lambda ^2 \mu ^2 V(\xi )-6\right)+8\right)\nonumber\\
 &+&16 \sin ^{11}(\xi ) Q(\xi )^2 \left(\cos (\xi ) V(\xi ) \left(\lambda ^2 \mu ^2 V(\xi )-6\right) -2 \sin (\xi ) V'(\xi )\right)\nonumber\\
 &+&2 \sin ^7(\xi ) Q(\xi )^4 \left(-3 \sin (\xi ) V'(\xi ) \left(\lambda ^2 \mu ^2 V(\xi )-8\right)-4 \cos (\xi ) V(\xi ) \left(\lambda ^2 \mu ^2 V(\xi )-12\right)\right)\nonumber\\
 &+&\sin ^3(\xi ) Q(\xi )^6 \left(\lambda ^2 \mu ^2 V(\xi )-4\right) \left(\sin (\xi ) V'(\xi ) \left(8-\lambda ^2 \mu ^2 V(\xi )\right)+4 \cos (\xi ) V(\xi ) \left(\lambda ^2 \mu ^2 V(\xi )+2\right)\right)\nonumber\\
 &+&8 \sin ^{15}(\xi ) \left(\sin (\xi ) V'(\xi )+4 \cos (\xi ) V(\xi )\right)=0.
\end{eqnarray}
With this cumbersome constraint equation, solutions for $V,Q,$ and $\xi$ are hard to get.

We could choose the potential to be zero, $V=0$, then we obtain the only nonzero component of the BPS Lagrangian density is
\begin{equation}
Q_{xy}=\pm\frac{4 f \lambda  \sin ^2(\xi )}{\left(f^2+1\right)^2},
\end{equation}
and the Bogomolny's equation is
\begin{equation}\label{BEqn5}
 f'(\theta ) g'(\varphi ) \sqrt{\frac{4}{\lambda ^2}+\xi '(r)^2}=\pm\frac{\left(f^2+1\right)^2 r^2 \sin (\theta ) \csc ^2(\xi ) \xi '(r)}{2 f \lambda }.
\end{equation}
Additional constraint equation from the BPS Lagrangian density is given by
\begin{equation}
 \xi '(r)^2 \cos (\xi (r)) \left(\lambda ^2 \xi '(r)^2+4\right)+\xi ''(r) \sin (\xi (r)) \left(\lambda ^2 \xi '(r)^2+6\right).
\end{equation}
This constraint equation is an equation for $\xi$, while the Bogomolny's equation (\ref{BEqn5}) is also implicitly an equation for $\xi$, where $f=\tan(\theta)$ and $g=n\varphi$,
\begin{equation}
 \xi '(r)= \frac{2 n}{\sqrt{r^4 \csc ^4(\xi )-\lambda ^2 n^2}},
\end{equation}
with $n=\pm1,\pm2,\pm3,\ldots$. We found that those equations are hardly to have common solutions and hence most likely lead to the vacuum solution as in the case of subsection \ref{third two-derivatives}.

\section{NEW SUBMODELS WITH THREE DERIVATIVE-TERMS}
In this section all submodels will have the same BPS Lagrangian density as in (\ref{gen BPS Lagrangian}), with definitions for $X,Y$ and $Z$ vary in each submodel and they will be written excplicitly. Following previous steps, we will solve $\mathcal{L}_{eff}-\mathcal{L}_{BPS}=0$ by considering it first as a quadratic equation of $X,Y$ and $Z$ subsequently.

\subsection{Submodel $\mathcal{L}^{(1)}_2+\mathcal{L}^{(2)}_2+\mathcal{L}_4$}\label{New Three Submodel 1}
In this submodel we extend the First BPS submodel in subsection \ref{The First BPS submodel} with the following effective Lagrangian density:
\begin{eqnarray}
 \mathcal{L}_{eff} &=&-\frac{4 \sin ^2(\xi )}{ r^2\left(f^2+1\right)^2}X^2-\frac{4 f^2 \sin ^2(\xi )}{ r^2\left(f^2+1\right)^2 \sin ^2(\theta )} Y^2 -\frac{ 16 f^2 \sin ^4(\xi )}{ r^4\left(f^2+1\right)^4 \sin ^2(\theta )}Z^2-\mu ^2 V,
\end{eqnarray}
with $X\equiv f'(\theta)\sqrt{1+\xi'(r)^2}, Y\equiv g'(\varphi)\sqrt{1+\xi'(r)^2}$ and $Z\equiv f'(\theta) g'(\varphi)$. It turns out the non-trivial solutions are possible only if $Q_0=Q_x=Q_y=Q_{xz}=Q_{zy}=0$ and
\begin{equation}\label{Qz and Qxy}
 Q_z=\pm \frac{8 f \mu  \sqrt{V} \sin ^2(\xi )}{\left(f^2+1\right)^2}, \qquad Q_{xy}=\pm \frac{8 f}{\left(f^2+1\right)^2 \csc ^2(\xi )}.
\end{equation}
The Bogomolny's equations are the same as the Bogomolny's equations (\ref{BEqn known 3}), where set $\mu=1$ for convenient, and (\ref{BEqn known 3-1}) in subsection \ref{known three-derivative}. The BPS Lagrangian density yields additional constraint equation which is simply written as
\begin{eqnarray} 
 &-&8 \left(V(\xi )^{3/2} \left(-10 \sin ^2(\xi ) V'(\xi )^2+8 \sin (\xi ) V(\xi ) \left(\sin (\xi ) V''(\xi )+\cos (\xi ) V'(\xi )\right)+32 V(\xi )^2\right)\right)\nonumber\\
 &&r^2 \sin ^2(\xi ) \left(4 \cot (\xi ) V(\xi )-V'(\xi )\right)^3 \left(  V'(\xi )+4   \cot (\xi ) V(\xi )+4 \cot (\xi ) \sqrt{V(\xi )}\right)=0.
\end{eqnarray}
Solving each term in the power $r$ of the constraint equation above give us $V= c_1 \sin^4\xi$. Unfortunately this will lead to vacuum solution since there is also implicit constraint from the Bogomolny's equation (\ref{BEqn known 3}), $r^2\sqrt{V}\csc^2\xi=\text{constant}$. As discussed in the subsection \ref{known three-derivative}, there might be other solutions for the potential that leads to non-trivial solutions. Taking $r^2\sqrt{V}\csc^2\xi=1$, the potential must satisfy the following constraint equation
\begin{eqnarray}
 &&-32 \sin (\xi ) V(\xi )^3 \left(2 \sin (\xi ) V''(\xi )+(5 \cos (\xi )+\cos (3 \xi )) V'(\xi )\right)+256 \left(\cos ^4(\xi )-1\right) V(\xi )^4\nonumber\\
 &&-\sin ^4(\xi ) V'(\xi )^4+80 \sin ^2(\xi ) V(\xi )^2 V'(\xi )^2+8 \sin ^3(\xi ) \cos (\xi ) V(\xi ) V'(\xi )^3+\nonumber\\
 &&\sqrt{V(\xi )} \left(12 \sin ^2(2 \xi ) V(\xi ) V'(\xi )^2-192 \sin (\xi ) \cos ^3(\xi ) V(\xi )^2 V'(\xi )-4 \sin ^3(\xi ) \cos (\xi ) V'(\xi )^3\right.\nonumber\\
 &&\qquad\qquad\left.+256 \cos ^4(\xi ) V(\xi )^3\right)=0.
\end{eqnarray}
Unfortunately, we are still unable to find a correct solution for the potential that produces non-trivial solution.

\subsection{Submodel $\mathcal{L}^{(1)}_2+\mathcal{L}^{(2)}_2+\mathcal{L}^{(1)}_4+\mathcal{L}^{(2)}_4+\mathcal{L}_6$}
Another possible extension of the First BPS submodel in subsection \ref{The First BPS submodel} is given by
\begin{eqnarray}
 \mathcal{L}_{eff} &=&-\frac{4 \sin ^2(\xi )}{ r^2\left(f^2+1\right)^2}X^2-\frac{4 f^2 \sin ^2(\xi )}{ r^2\left(f^2+1\right)^2 \sin ^2(\theta )} Y^2 -\frac{ 4\lambda^2 f^2 \sin ^4(\xi )}{ r^4\left(f^2+1\right)^4 \sin ^2(\theta )}Z^2-\mu ^2 V,
\end{eqnarray}
with  $X\equiv f'(\theta)\sqrt{1+\xi'(r)^2}, Y\equiv g'(\varphi)\sqrt{1+\xi'(r)^2}$ and $Z\equiv f'(\theta) g'(\varphi) \xi'(r)$. This effective Lagrangian density is similar with the one in subsection \ref{New Three Submodel 1}, by shifting $Z\to {\lambda\over 2}Z$. We will then obtain the Bogomolny's equations (\ref{BEqn known 3-1}) and (\ref{BEqn known one derivative}). With the same nonzero $Q_z$ and $Q_{xy}$ as in (\ref{Qz and Qxy}), the additional constraint equation from the BPS Lagrangian density is given by (\ref{BEqn First BPS submodel}). As mentioned in subsection \ref{The First BPS submodel}, its solutions are compactons. As a an example we pick the compacton solution given in~\cite{Adam:2017pdh},
\begin{equation}
\xi(r)=\begin{cases}
\pi -r,\qquad 0\leq r \leq {\pi}\\
0,\qquad\qquad r> {\pi}
\end{cases}.
\end{equation}
However, the Bogomolny's equation (\ref{BEqn known one derivative}) implicitly has an equation for $\xi$,
\begin{equation}
 \xi'(r)=c_1~ r^2\sqrt{V} \csc(\xi)^2,
\end{equation}
with $c_1$ is a constant. Using the compacton solution above, the corresponding potential is
\begin{equation}
 V=\frac{\sin ^4(\xi )}{c_1^2 (\pi -\xi )^4}.
\end{equation}
Within the range of $0\leq\xi\leq\pi$, this potential has a minimum at $\xi=0$.

\section{Conclusions and Remarks}

We have used the BPS Lagrangian method to derive Bogomolny's equations in BPS submodels of the generalized Skyrme model with a particular ansatz (\ref{eq:ansatz}) in spherical coordinates. We argued that in the BPS Lagrangian method the general BPS Lagrangian density for the generalized Skyrme model is given by (\ref{gen BPS Lagrangian}) while the effective Lagrangian density consist at most three derivative-terms effectively. We then derived rigorously the Bogomolny's equations for all the known BPS submodels of the generalized Skyrme model. There was also an additional constraint equation depending on the explicit form of the BPS Lagrangian density. The constraint equation was merely non-trivial Euler-Lagrange equation of the BPS Lagrangian density as discussed in~\cite{Atmaja:2018ddi}. Some of the Bogomolny's equations, along with the constraint equation, are quite different from the ones found in the literature. However, the solutions found in the literature are indeed solutions of these Bogomolny's equations and furthermore there might be also other possible solutions as we have shown explicitly in subsections \ref{The First BPS submodel} and \ref{known three-derivative}.

We also gave examples for possible new BPS submodels of the generalized Skyrme model starting from BPS submodel with one derivative-term up to three derivative-terms. We derived and tried to solve the Bogomolny's equations, along with the constraint equation, of new BPS submodels. Some of the solutions were relatively new while the other ones were hard to solve due to complicated constraint equation such as submodels in subsections \ref{New BPS submodel II-D} and \ref{New Three Submodel 1}.

Here we have used a particular ansatz (\ref{eq:ansatz}), similar to the natural (hedgehog) ansatz in~\cite{Adam:2010fg}, in deriving the Bogomolny equations of the BPS submodels in the generalized Skyrme model. We can try to use a different ansatz while employing the BPS Lagrangian method and then we might obtain different possible BPS submodels other than the ones obtained here. We could also loosen up the requirement for all the $Q$'s in the BPS Lagrangian density (\ref{gen BPS Lagrangian}) by allowing them to depend on the spatial coordinates explicitly. However, this requirement will make the analysis becoming more involved.

The BPS Lagrangian method opens many possibilities in deriving Bogomolny equations. It allows terms to be included in the BPS Lagrangian density that are not boundary terms as shown here, while other methods such as the FOEL method in~\cite{Adam:2016ipc} only allows to add the boundary terms in the Lagrangian, which corresponds to the BPS Lagrangian density containing only boundary terms. So far the BPS Lagrangian method works in the effective Lagrangian density description and it would be nice to have this method to works in more general description, without a priori imposing an ansatz, such that we could connect the resulting Bogomolny equations with the BPS equations in the context of supersymmetric theories.

\section{ACKNOWLEDGEMENT}
We would like to thank C. Adam for discussions. IP is greatful to Research Center for Physics-LIPI for financial support under the Reseach Assistant scheme. ANA and BEG acknowledge the Abdus Salam ICTP for Associateships 2019 and for warmest hospitality. The work of ANA and BEG is supported by PDUPT Kemenristek 2020.

\appendix

\bibliography{referensi}

\end{document}